\documentclass[]{article}
\usepackage[utf8x]{inputenc}
\usepackage{amsmath,amsfonts,amsthm}
\usepackage{graphicx}
\usepackage[affil-it]{authblk}
\usepackage[a4paper,margin=1in]{geometry}
\numberwithin{equation}{section}
\allowdisplaybreaks

\newmuskip\pFqskip
\mathchardef\pFcomma=\mathcode`, 

\newcommand*\pFq[5]{%
  \begingroup
  \begingroup\lccode`~=`,
    \lowercase{\endgroup\def~}{\pFcomma\mkern\pFqskip}%
  \mathcode`,=\string"8000
  {}_{#1}F_{#2}\biggl(\genfrac..{0pt}{}{#3}{#4} \Big| #5\biggr)%
  \endgroup
}

\newcommand*{\Scale}[2][4]{\scalebox{#1}{$#2$}}%

\title{A superintegrable model with reflections on $S^3$ \\ and the rank two Bannai-Ito algebra}
\author{Hendrik De Bie}
\affil{Department of Mathematical Analysis, Faculty of Engineering and Architecture, Ghent University, \protect\\Galglaan 2, 9000 Ghent, Belgium \authorcr E-mail address: Hendrik.DeBie@UGent.be\protect\\ \ } 
\author{Vincent X. Genest}
\affil{Department of Mathematics, Massachusetts Institute of Technology, 77 Massachusetts Ave., \protect\\Cambridge, MA 02139, USA \authorcr E-mail address: vxgenest@mit.edu \protect\\ \ }
\author{Jean-Michel Lemay}
\author{Luc Vinet}
\affil{Centre de Recherches Math\'ematiques, Universit\'e de Montr\'eal, C.P. 6128, Succ. Centre-ville,\protect\\ Montr\'eal, QC, Canada, H3C 3J7 \authorcr E-mail address: jean-michel.lemay.1@umontreal.ca, vinet@crm.umontreal.ca}
\date{}

\begin{document}
\maketitle

\begin{abstract}

\noindent A quantum superintegrable model with reflections on the three-sphere is presented. Its symmetry algebra is identified with the rank-two Bannai-Ito algebra. It is shown that the Hamiltonian of the system can be constructed from the tensor product of four representations of the superalgebra $\mathfrak{osp}(1|2)$ and that the superintegrability is naturally understood in that setting. The exact separated solutions are obtained through the Fischer decomposition and a Cauchy-Kovalevskaia extension theorem.\\

\noindent This paper is dedicated with admiration and gratitude to Ji\v{r}\'{\i} Patera and Pavel Winternitz on the occasion of their 80th birthdays.
 
\end{abstract}

\section{Introduction}

 Superintegrability shares an intimate connection with exact solvability. For classical systems, this connection is fully understood while it remains an empirical observation for general quantum systems. The study of superintegrable models has proved fruitful in understanding symmetries and their algebraic description, and has also contributed to the theory of special functions. A quantum system in $n$ dimensions with Hamiltonian $H$ is said to be maximally superintegrable if it possesses $2n-1$ algebraically independent constants of motion $c_1, c_2,\ldots, c_{2n-1}$  commuting with $H$, that is $[H,c_i]=0$ for $i=1,\ldots,2n-1$, where one of these constants is the Hamiltonian itself. Such a system is further said to be superintegrable of order $l$ if the maximum order in momenta of the constants of motion (except $H$) is $l$.

 One of the important quantum superintegrable models is the so-called generic three-parameter system on the two-sphere \cite{2013_Miller&Post&Winternitz_SI&apps}, whose symmetries generate the Racah algebra which characterizes the Wilson and Racah polynomials sitting atop the Askey scheme \cite{Askey_scheme}. All two-dimensional second order superintegrable models of the form $H=\Delta+V$ where $\Delta$ denotes the Laplace-Beltrami operator have been classified \cite{2013_Miller&Post&Winternitz_SI&apps} and can be obtained from the generic three-parameter model through contractions and specializations \cite{2013_Kalnins&Miller&Post_2DContractions}. A similar model with four parameters defined on the three-sphere has also been introduced and its connection to bivariate Wilson and Racah polynomials has been established \cite{2013_Kalnins&Miller&Post_2varWilson}.

  Recently, superintegrable models defined by Hamiltonians involving reflection operators have been the subject of several investigations \cite{dunklosc,dunklosc2,dunklosc_anis, dunklosc3D, 2011_Post&Vinet&Zhedanov_SusyQM}. One of the interesting features of these models is their connection to less known bispectral orthogonal polynomials referred to as $-1$ polynomials. Many efforts have been deployed to characterize these polynomials, which can be organized in a tableau similar to the Askey one \cite{family, bigjac, bochner, dual, BI, CBI, ChiharaP}. Of particular relevance to the present paper is the Laplace-Dunkl equation on the two-sphere studied in \cite{laplacedunkl, BIalgebra_2sphere}, which has the rank-one Bannai-Ito algebra as its symmetry algebra \cite{BI}. This Bannai-Ito algebra encodes the bispectrality of the Bannai-Ito polynomials which depend on four parameters and stand at the highest level of the hierarchy of $-1$ orthogonal polynomials. As such, this Laplace-Dunkl system on the two sphere can be thought of as a generalization with reflection operators of the generic three-parameter model (without reflections) on the two-sphere which is recovered when wavefunctions with definite parities are considered. The goal of this paper is to introduce a novel quantum superintegrable model with reflections on the three-sphere which similarly embodies the generic four-parameter model introduced and studied in \cite{2013_Kalnins&Miller&Post_2varWilson}.  

  The paper is divided as follows. In section 2, we introduce a superintegrable model with four-parameters on the three-sphere and exhibit its symmetries explicitly. In section 3, it is shown how the Hamiltonian of the model can be constructed from four realizations of the superalgebra $\mathfrak{osp}(1|2)$. Moreover, the symmetry algebra is characterized and is seen to correspond to a rank-two generalization of the Bannai-Ito algebra. In section 4, the structure of the space of polynomial solutions is exhibited using a Fischer decomposition and an explicit basis for the eigenfunctions is constructed with the help of a Cauchy-Kovalevskaia extension theorem. Some concluding remarks are offered in section 6.

\section{A superintegrable model on $S^3$}

  Let $s_1,s_2,s_3, s_4$ be the Cartesian coordinates of a four-dimensional Euclidian space and take the restriction to the embedded three-sphere: $s_1^2+s_2^2+s_3^2+s_4^2=1$. Consider the system with four parameters $\mu_1, \mu_2, \mu_3, \mu_4$ with $\mu_i\geq 0$ for $i=1,2,3,4$ governed by the Hamiltonian  
  \begin{align} \label{H}
  H = \sum_{1\le i<j\le4} J_{ij}^2 +\sum_{i=1}^4 \frac{\mu_i}{s_i^2}(\mu_i - R_i),
  \end{align}
  where 
  \begin{align}
  J_{ij} = \frac{1}{i}(s_i \partial_{s_j} - s_j \partial_{s_i}), \qquad \qquad  R_i f(s_i) = f(-s_i),
  \end{align}
  are the angular momentum operators and reflection operators, respectively. The six quantities
  \begin{align}
  L_{jk} = \Bigg(\frac{1}{2}+\mu_jR_j+\mu_kR_k+\left(iJ_{jk}+\mu_j\frac{s_k}{s_j}R_j-\mu_k\frac{s_j}{s_k}R_k\right)\prod_{l=j+1}^{k}R_l\Bigg)R_jR_k , \qquad 1\le j<k\le4,
  \end{align}
  can easily be verified to commute with $H$ on the 3-sphere and are thus conserved. It can be shown that any four of the $L_{jk}$ are algebraically independent. Hence $H$ defines a maximally superintegrable system of first order. There are also four more conserved quantities of the form
  \begin{align}
  M_A = \Bigg(1+\sum_{i\in A}\mu_iR_i + \sum_{\substack{j<k\\ j,k\in A}}\left(iJ_{jk}+s_k\frac{\mu_j}{s_j}R_j-s_j\frac{\mu_k}{s_k}R_k\right)\prod_{l=j+1}^{k}R_l \Bigg) \prod_{i\in A} R_i,
  \end{align}
  where $A=\{1,2,3\},\{1,2,4\},\{1,3,4\}$ or $\{2,3,4\}$. Furthermore, a direct computation yields
  \begin{align}
  [H,R_i]=0, \qquad i=1,2,3,4.
  \end{align}
  The reflections are thus discrete symmetries of the system. 

\section{Algebraic construction from $\mathfrak{osp}(1|2)$}  

  The superalgebra $\mathfrak{osp}(1|2)$ can be presented with five generators $x, D, E, |x|^2$ and $D^2$ with the following defining relations:
  \begin{align} \label{osprelns}
  \begin{aligned}
  &\{x,x\}=2|x|^2,  &&\{D,D\}=2D^2, \\
  &\{x,D\}=2E,    &&[D,E]=D, \\
  &[D,|x|^2]=2x,  &&[E,x]=x, \\
  &[D^2,x]=2D,    &&[D^2,E]=2D^2,\\
  &[D^2,|x|^2]=4E, &&[E,|x|^2]=2|x|^2,
  \end{aligned}
  \end{align}
  where $[a,b]=ab-ba$ is the commutator and $\{a,b\}=ab+ba$ is the anti-commutator. One can realize four mutually commuting copies of this superalgebra by taking
  \begin{align} \label{irealization}
  \begin{aligned}
  &D_i = \partial_{s_i} - \frac{\mu_i}{s_i}R_i,  &&D^2_i= D_iD_i, \\
  &x_i=s_i, &&|x_i|^2 = s_i^2, \\
  &E_i=s_i\partial_{s_i}+\frac{1}{2},
  \end{aligned}
  \end{align}
  where $i=1,2,3,4$. Each superalgebra possesses a sCasimir element given by
  \begin{align}
  S_i = \frac{1}{2}([D_i,x_i]-1),
  \end{align}
  which anticommutes with the odd generators
  \begin{align}
  \{S_i,D_i\}=\{S_i,x_i\}=0,
  \end{align}
  and thus commutes with the even generators
  \begin{align}
  [S_i,E_i]=[S_i,|x_i|^2]=[S_i,D_i^2]=0.
  \end{align}
  It is immediate to verify that in the realization \eqref{irealization}, the reflection $R_i$ verifies the same commutation relations as the sCasimir
  \begin{align}
  [R_i,E_i]=[R_i,|x_i|^2]=[R_i,D_i^2]=\{R_i,D_i\}=\{R_i,x_i\}=0.
  \end{align}
  This implies that one can construct a Casimir operator of the form
  \begin{align}
  Q_i = S_iR_i.
  \end{align}
  It is straightforward to verify that $Q_i$ indeed commutes with every generator. These four realizations of $\mathfrak{osp}(1|2)$ can act as building blocks for many other realizations. Let $[n]=\{1,2,\dots,n\}$ and $A\subset [4]$. The operators given by
  \begin{align} \label{Arealization}
  \begin{aligned}
  &D_A = \sum_{i\in A}\Big( D_i\prod_{j=i+1}^{\sup A}R_j\Big), &&D^2_A= D_AD_A, \\
  &x_A=\sum_{i\in A}\Big(s_i\prod_{j=i+1}^{\sup A}R_j\Big),    &&|x_A|^2 = \sum_{i\in A}s_i^2, \\
  &E_A=\sum_{i\in A}E_i,
  \end{aligned}
  \end{align}
  verify the commutation relations \eqref{osprelns} for any $A\subset [4]$ and thus form new realizations of $\mathfrak{osp}(1|2)$. These result from the repeated application of the coproduct of $\mathfrak{osp}(1|2)$ (see \cite{Embeddings}). Moreover, for any $A$ the sCasimir and the Casimir operators are also similarly defined:
  \begin{align} \label{Arealization2}
  S_A = \frac{1}{2}([D_A,x_A]-1), \quad\qquad Q_A = S_A\prod_{i\in A}R_i.
  \end{align}
  One can directly check that
  \begin{align}
  Q_i = \mu_i, \qquad Q_{jk}=L_{jk}, \qquad Q_B = M_B,
  \end{align}
  where $Q_{jk}$ denotes $Q_A$ with $A=\{j,k\}$ and $B$ is any 3-subset of $[4]$. Another explicit computation gives
  \begin{align} \label{HsCasimir}
  S_{[4]}^2-S_{[4]}-\tfrac{3}{4} = \sum_{1\le i<j\le4} J_{ij}^2 +(s_1^2+s_2^2+s_3^2+s_4^2)\sum_{i=1}^4 \frac{\mu_i}{s_i^2}(\mu_i - R_i).
  \end{align}
  However, since $|x_{[4]}|^2=s_1^2+s_2^2+s_3^2+s_4^2$ commutes with $S_{[4]}$ and all the Casimirs, it is central in the algebra generated by the Casimirs and can thus be treated as a constant. Taking $|x_{[4]}|^2=1$, it is straightforward by comparing \eqref{HsCasimir} and \eqref{H} that
  \begin{align} \label{algebraicH}
  S_{[4]}^2-S_{[4]}-\tfrac{3}{4} = H.
  \end{align}
  Hence, a quadratic combination of the sCasimir of four copies of $\mathfrak{osp}(1|2)$ yields the Hamiltonian of the superintegrable model presented in section 1 and the intermediate Casimirs are its symmetries. Indeed, it can be checked that $[Q_A,H]=0$ for $A \subset [4]$. The symmetry algebra has the following structure relations 
  \begin{align} \label{BIrank2}
  \{ Q_A,Q_B\} = Q_{(A\cup B)\setminus (A\cap B)}+2Q_{A\cap B}Q_{A\cup B} + 2Q_{A\setminus (A\cap B)}Q_{B\setminus (A\cap B)},
  \end{align}
  where $A,B\subset[4]$ and $Q_\emptyset = -1/2$ as prescribed by the definitions \eqref{Arealization} and \eqref{Arealization2}. This algebra has already been studied in \cite{higherrankBIalgebra} and is interpreted as a rank 2 Bannai-Ito algebra. To see this, we remark that the Casimirs with $A \subset [3]$ generate the (rank 1) Bannai-Ito algebra. Let $K_1 = Q_{12}, K_2 = Q_{23}$ and $K_3 = Q_{13}$. The recurrence relations \eqref{BIrank2} can then be rewritten as
  \begin{align}
   \{K_1,K_2\}=K_3+\omega_3, \quad \{K_2,K_3\}=K_1+\omega_1, \quad \{K_3,K_1\}=K_2+\omega_2, 
  \end{align}
  where $\omega_1, \omega_2, \omega_3$ are central elements given by
  \begin{align}
   \omega_1 =2Q_3Q_{123}+2Q_1Q_2, \quad \omega_2 = 2Q_1Q_{123}+2Q_2Q_3, \quad \omega_3 = 2Q_2Q_{123}+2Q_1Q_3.
  \end{align}
  This corresponds to the Bannai-Ito algebra introduced in \cite{BI} which appears in a corresponding superintegrable model with reflections on $S^2$ as its symmetry algebra \cite{laplacedunkl}.  

\section{Wavefunctions}
  To obtain the solutions to the equation $H\psi = \lambda\psi$ let us first introduce the gauge transformation
  \begin{align}
  z\to \tilde{z}\equiv G(\vec{s})^{-1} z G(\vec{s}), \qquad G(\vec{s})=\prod_{i=1}^4 |s_i|^{\mu_i},
  \end{align}
  where $z$ is any operator and $\vec{s}\equiv(s_1,s_2,s_3,s_4)$. Under this transformation, the generators of $\mathfrak{osp}(1|2)$ in the realization \eqref{irealization} become
  \begin{align} \label{tildeRealization}
  \begin{aligned}
   &\tilde{D_i} = \partial_{s_i} + \frac{\mu_i}{s_i}(1-R_i),\quad &&\tilde{D}^2_i= \tilde{D}_i\tilde{D}_i,  \\
   &\tilde{x_i} = x_i = s_i, && |\tilde{x}_i|^2=s_i^2,\\
   &\tilde{E_i} = s_i\partial_{s_i} +\gamma_i, &&\tilde{R_i} = R_i,\\
   &\tilde{S_i}=-\mu_iR_i, &&\tilde{Q_i} = \mu_i ,
  \end{aligned}
  \end{align}
  where
  \begin{align}
  \gamma_A = \sum_{i\in A}(\mu_i + \tfrac{1}{2}). 
  \end{align}
  These operators also verify \eqref{osprelns} and correspond to the realization of $\mathfrak{osp}(1|2)$ (or equivalently $sl_{-1}(2)$) arising in the one-dimensional parabose oscillator \cite{dunklosc}. Furthermore, the construction \eqref{Arealization} can be reproduced with this transformed realization to obtain operators of the form $\tilde{D}_A, \tilde{x}_A, \tilde{E}_A,\tilde{S}_A$ and $\tilde{Q}_A$ and is trivially seen to be equivalent to the gauge transformation of the corresponding operators. Hence, we can obtain eigenvalues and eigenfunctions of $H$ by finding eigenfunctions of $\tilde{S}_{[4]}$. Note that since $\tilde{S}_{[4]}$ commutes with $P=R_1R_2R_3R_4$, this is equivalent to finding eigenfunctions of $\tilde{Q}_{[4]}$. 

  We thus aim to obtain polynomial eigenfunctions of $\tilde{S}_{[4]}$. To do so, let us first introduce $\mathcal{P}_m(\mathbb{R}^n)$, the space of homogeneous polynomials of degree $m$ in the variables $s_1,s_2,\dots,s_n$. We define $\mathcal{K}_m(\mathbb{R}^n)$ the kernel space of degree $m$ as 
  \begin{align}
   \mathcal{K}_m(\mathbb{R}^n) = \ker \tilde{D}_{[n]} \cap \mathcal{P}_m(\mathbb{R}^n).
  \end{align}
  When $n=4$, this is an eigenspace of $\tilde{S}_{[4]}$. Indeed, take $\psi_m \in \mathcal{K}_m(\mathbb{R}^4)$ and compute
  \begin{align*}
   \tilde{S}_{[4]}\tilde{\psi}_m &= \tfrac{1}{2}(\tilde{D}_{[4]}\tilde{x}_{[4]}-\tilde{x}_{[4]}\tilde{D}_{[4]}-1)\tilde{\psi}_m = \tfrac{1}{2}(\tilde{D}_{[4]}\tilde{x}_{[4]}-1)\tilde{\psi}_m \\
    &=\tfrac{1}{2}(\tilde{D}_{[4]}\tilde{x}_{[4]}+\tilde{x}_{[4]}\tilde{D}_{[4]}-1)\tilde{\psi}_m = \tfrac{1}{2}(\{\tilde{x}_{[4]},\tilde{D}_{[4]}\}-1)\tilde{\psi}_m \\
    &=\tfrac{1}{2}(2\tilde{E}_{[4]}-1)\tilde{\psi}_m = \left[\sum_{i=1}^4 s_i\partial_{s_i} +\gamma_{[4]}-\tfrac{1}{2}\right]\tilde{\psi}_m,
  \end{align*}
  where we used the property $\tilde{D}_{[4]}\tilde{\psi}_m=0$ and the commutation relations \eqref{osprelns}. Since $\tilde{\psi}_m$ is a homogeneous polynomial of degree $m$, it is an eigenfunction of the Euler operator : $\sum_{i=1}^4 s_i\partial_{s_i} \tilde{\psi}_m = m \tilde{\psi}_m $. This implies 
  \begin{align} \label{eigenv}
   \tilde{S}_{[4]}\tilde{\psi}_m = (m+\gamma_{[4]}-\tfrac{1}{2})\tilde{\psi}_m
  \end{align}
   and shows that $\mathcal{K}_m(\mathbb{R}^4)$ is an eigenspace of $\tilde{S}_{[4]}$. We use two results in order to construct explicitly the eigenfunctions. First, the space of homogeneous polynomials $\mathcal{P}_m(\mathbb{R}^n)$ admits a decomposition in terms of the kernel spaces. This is called the Fischer decomposition and can be cast as 
   \begin{align} \label{Fischer}
    \mathcal{P}_m(\mathbb{R}^n) = \bigoplus_{j=0}^{m} \tilde{x}_{[n]}^j \mathcal{K}_{m-j}(\mathbb{R}^n).
   \end{align}
   Second, we use the Cauchy-Kovalevskaia isomorphism (CK-map) between the space of $m$-homogeneous polynomials in $n-1$ variables and the kernel space of degree $m$ in $n$ variables :
   \begin{align} \label{CKmap}
    {\bf CK}_{s_n}^{\mu_n} : \mathcal{P}_m(\mathbb{R}^{n-1}) \to \mathcal{K}_{m}(\mathbb{R}^n).
   \end{align}
  One can compute the CK-map explicitly. To compute ${\bf CK}_{s_4}^{\mu_4}$, take $p(s_1,s_2,s_3)\in \mathcal{P}_m(\mathbb{R}^3)$ and let
  \begin{align}
   {\bf CK}_{s_4}^{\mu_4}[p(s_1,s_2,s_3)]=\sum_{\alpha=0}^{m}s_4^\alpha p_\alpha(s_1,s_2,s_3),
  \end{align}
  where $p_\alpha(s_1,s_2,s_3)\in \mathcal{P}_{m-\alpha}(\mathbb{R}^3)$ and $p_0(s_1,s_2,s_3)\equiv p(s_1,s_2,s_3)$. Demand that 
  \begin{align}
  \tilde{D}_{[4]}\sum_{\alpha=0}^{m}s_4^\alpha p_\alpha(s_1,s_2,s_3)=0 
  \end{align}
  to fix and compute the coefficients $p_\alpha(s_1,s_2,s_3)$. A straightforward calculation yields
  \begin{align}
   {\bf CK}_{s_4}^{\mu_4}= \sum_{i=0}^{\infty} \frac{(-1)^i(s_4)^{2i}}{i!(\gamma_4)_i(2)^{2i}}\tilde{D}_{[4]}^{2i} + \sum_{i=0}^{\infty} \frac{(-1)^{i+1}(s_4)^{2i+1}}{i!(\gamma_4)_{i+1}(2)^{2i+1}}\tilde{D}_{[4]}^{2i+1},
  \end{align}
  where $(a)_i = a(a+1)\dots(a+n-1)$ denotes the Pochhammer symbol. Similarly, one obtains
  \begin{align}
   {\bf CK}_{s_n}^{\mu_n}= \sum_{i=0}^{\infty} \frac{(-1)^i(s_n)^{2i}}{i!(\gamma_n)_i(2)^{2i}}\tilde{D}_{[n]}^{2i} + \sum_{i=0}^{\infty} \frac{(-1)^{i+1}(s_n)^{2i+1}}{i!(\gamma_n)_{i+1}(2)^{2i+1}}\tilde{D}_{[n]}^{2i+1},
  \end{align}
  for $n=2,3,4$. Now, iterating the Fischer decomposition \eqref{Fischer} and the CK-map \eqref{CKmap}, the eigenspace $\mathcal{K}_m(\mathbb{R}^4)$ can be expressed as
   \begin{align}
    \mathcal{K}_m(\mathbb{R}^4) \cong {\bf CK}_{s_4}^{\mu_4}\Big[\bigoplus_{j_2=0}^{m} \tilde{x}_{[3]}^{m-j_2}{\bf CK}_{s_3}^{\mu_3}\Big[\bigoplus_{j_1=0}^{j_2} \tilde{x}_{[2]}^{j_2-j_1}{\bf CK}_{s_2}^{\mu_2}\left[\mathcal{P}_{j_1}(\mathbb{R})\right]\Big]\Big].
   \end{align}
   This means that we can explicitly construct a basis of eigenfunctions $\{\tilde{\psi}_{j_1,j_2,j_3}^{(m)}(\vec{s})\}_{j_1+j_2+j_3=m}$ of $\mathcal{K}_m(\mathbb{R}^4)$ with
   \begin{align}
    \tilde{\psi}_{j_1,j_2,j_3}^{(m)}(\vec{s}) = {\bf CK}_{s_4}^{\mu_4}\big[ \tilde{x}_{[3]}^{j_3}{\bf CK}_{s_3}^{\mu_3}\big[\tilde{x}_{[2]}^{j_2}{\bf CK}_{s_2}^{\mu_2}[s_1^{j_1}]\big]\big].
   \end{align}
   This calculation can be carried straightforwardly with the help of the identities with $\tilde{\psi}_m \in \mathcal{K}_m(\mathbb{R}^n)$ 
   \begin{align}
   \begin{aligned}
   \tilde{D}_{[n]}^{2\alpha}\tilde{x}_{[n]}^{2\beta}\tilde{\psi}_m &= 2^{2\alpha}(-\beta)_\alpha(1-m-\beta-\gamma_{[n]})_\alpha x_{[n]}^{2\beta-2\alpha}\tilde{\psi}_m, \\
   \tilde{D}_{[n]}^{2\alpha+1}\tilde{x}_{[n]}^{2\beta}\tilde{\psi}_m &= 2^{2\alpha}\beta(1-\beta)_\alpha(1-m-\beta-\gamma_{[n]})_\alpha x_{[n]}^{2\beta-2\alpha-1} \tilde{\psi}_m, \\
   \tilde{D}_{[n]}^{2\alpha}\tilde{x}_{[n]}^{2\beta+1}\tilde{\psi}_m &= 2^{2\alpha}(-\beta)_\alpha(-m-\beta-\gamma_{[n]})_\alpha x_{[n]}^{2\beta+1-2\alpha}\tilde{\psi}_m, \\
   \tilde{D}_{[n]}^{2\alpha+1}\tilde{x}_{[n]}^{2\beta+1}\tilde{\psi}_m &= 2^{2\alpha+1}(-\beta)_\alpha(m+\beta+\gamma_{[n]})(1-m-\beta-\gamma_{[n]})_\alpha x_{[n]}^{2\beta-2\alpha} \tilde{\psi}, 
    \end{aligned}
   \end{align}
   which follows from \eqref{osprelns}. The result can be presented in terms of the Jacobi polynomials $P_n^{(\alpha,\beta)}(x)$, defined as \cite{Askey_scheme}
   \begin{align}
    P_n^{(\alpha,\beta)}(x) = \frac{(\alpha+1)_n}{n!} \pFq{2}{1}{-n,n+\alpha+\beta+1}{\alpha+1}{\frac{1-x}{2}},
   \end{align}
   with the help of the identity :
   \begin{align}
   (x+y)^n P_n^{(\alpha,\beta)}\left(\frac{x-y}{x+y}\right) = \frac{(\alpha+1)_n}{n!}x^n\pFq{2}{1}{-n,-n-\beta}{\alpha+1}{-\frac{y}{x}}. 
   \end{align}
   One obtains
   \begin{align}
    \tilde{\psi}_{j_1,j_2,j_3}^{(m)}(\vec{s}) = {\bf P}_{j_1,j_2,j_3}(\vec{s}){\bf Q}_{j_1,j_2}(s_1,s_2,s_3)R_{j_1}(s_1,s_2),
   \end{align}
   where
   \begin{align*}
   \begin{aligned}
    &{\bf P}_{j_1,j_2,j_3}(\vec{s}) = \frac{c!}{(\gamma_4)_c}(s_1^2+s_2^2+s_3^2+s_4^2)^c \\ 
    &\times \begin{cases}
    \begin{aligned}
    P_c^{(\gamma_4-1,j_1+j_2+\gamma_{[3]}-1)}\Scale[0.9]{\left(\frac{s_1^2+s_2^2+s_3^2-s_4^2}{s_1^2+s_2^2+s_3^2+s_4^2}\right)}
    - \Scale[1]{\frac{s_4 \tilde{x}_{[3]}}{s_1^2+s_2^2+s_3^2+s_4^2}} P_{c-1}^{(\gamma_4,j_1+j_2+\gamma_{[3]})}\Scale[0.9]{\left(\frac{s_1^2+s_2^2+s_3^2-s_4^2}{s_1^2+s_2^2+s_3^2+s_4^2}\right)}
    \end{aligned}
     &\Scale[1]{\text{if $j_3=2c$,}} \\[1em]
    \begin{aligned}
    \tilde{x}_{[3]} P_c^{(\gamma_4-1,j_1+j_2+\gamma_{[3]})}\Scale[0.9]{\left(\frac{s_1^2+s_2^2+s_3^2-s_4^2}{s_1^2+s_2^2+s_3^2+s_4^2}\right)}
    - \Scale[1]{s_3 \frac{j_1+j_2+c+\gamma_{[3]}}{c+\gamma_4}} P_c^{(\gamma_4,j_1+j_2+\gamma_{[3]}-1)}\Scale[0.9]{\left(\frac{s_1^2+s_2^2+s_3^2-s_4^2}{s_1^2+s_2^2+s_3^2+s_4^2}\right)}
    \end{aligned}
     &\Scale[1]{\text{if $j_3=2c+1$,}} 
    \end{cases}
   \end{aligned}
   \end{align*}
   \begin{align*}
   \begin{aligned}
    {\bf Q}_{j_1,j_2}&(s_1,s_2,s_3) = \frac{b!}{(\gamma_3)_b}(s_1^2+s_2^2+s_3^2)^b \\ 
    &\times \begin{cases}
    \begin{aligned}
    P_b^{(\gamma_3-1,j_1+\gamma_{[2]}-1)}\Scale[1]{\left(\frac{s_1^2+s_2^2-s_3^2}{s_1^2+s_2^2+s_3^2}\right)}
    - \Scale[1]{\frac{s_3 \tilde{x}_{[2]}}{s_1^2+s_2^2+s_3^2}} P_{b-1}^{(\gamma_3,j_1+\gamma_{[2]})}\Scale[1]{\left(\frac{s_1^2+s_2^2-s_3^2}{s_1^2+s_2^2+s_3^2}\right)}
    \end{aligned}
     &\Scale[1]{\text{if $j_2=2b$,}} \\[1em]
    \begin{aligned}
    \tilde{x}_{[2]} P_b^{(\gamma_3-1,j_1+\gamma_{[2]})}\Scale[1]{\left(\frac{s_1^2+s_2^2-s_3^2}{s_1^2+s_2^2+s_3^2}\right)}
    - \Scale[1]{s_2 \frac{j_1+b+\gamma_{[2]}}{b+\gamma_3}} P_b^{(\gamma_3,j_1+\gamma_{[2]}-1)}\Scale[1]{\left(\frac{s_1^2+s_2^2-s_3^2}{s_1^2+s_2^2+s_3^2}\right)}
    \end{aligned}
     &\Scale[1]{\text{if $j_2=2b+1$,}} 
    \end{cases}
   \end{aligned}
   \end{align*} 
   \begin{align*}
   \begin{aligned}
    R_{j_1}(s_1,s_2) = \frac{a!}{(\gamma_2)_a}(s_1^2+s_2^2)^a 
    \times \begin{cases}
    \begin{aligned}
    P_a^{(\gamma_2-1,\gamma_1-1)}\Scale[1]{\left(\frac{s_1^2-s_2^2}{s_1^2+s_2^2}\right)}
    - \Scale[1]{\frac{s_1s_2}{s_1^2+s_2^2}} P_{a-1}^{(\gamma_2,\gamma_1)}\Scale[1]{\left(\frac{s_1^2-s_2^2}{s_1^2+s_2^2}\right)}
    \end{aligned}
     &\Scale[1]{\text{if $j_1=2a$,}} \\[1em]
    \begin{aligned}
    s_1 P_a^{(\gamma_2-1,\gamma_1)}\Scale[1]{\left(\frac{s_1^2-s_2^2}{s_1^2+s_2^2}\right)}
    - \Scale[1]{s_2 \frac{a+\gamma_1}{a+\gamma_2}} P_a^{(\gamma_2,\gamma_1-1)}\Scale[1]{\left(\frac{s_1^2-s_2^2}{s_1^2+s_2^2}\right)}
    \end{aligned}
     &\Scale[1]{\text{if $j_1=2a+1$.}} 
    \end{cases}
   \end{aligned}
   \end{align*} 
  Note that the expressions for ${\bf P}_{j_1,j_2,j_3}(s_1,s_2,s_3,s_4)$ and ${\bf Q}_{j_1,j_2}(s_1,s_2,s_3)$ contain the operators $\tilde{x}_{[3]}$ and $\tilde{x}_{[2]}$ respectively. Recalling the expressions \eqref{tildeRealization} and \eqref{Arealization}, it can be seen that these operators only contain variables $s_i$ and reflection operators $R_i$. These reflections conveniently account for signs occuring in the solutions without having to give a different expression for every parity combination of the parameters $j_1,j_2$ and $j_3$.   

  By effecting the reverse gauge transformation, we thus obtain a basis for the eigenspace of the operator $S_{[4]}$ given by
  \begin{align}
   \psi_{j_1,j_2,j_3}^{(m)}(\vec{s}) = \tilde{\psi}_{j_1,j_2,j_3}^{(m)}(\vec{s})G(\vec{s}), 
  \end{align}
  where $m=0,1,\dots$ and $j_1+j_2+j_3=m$. With the help of \eqref{eigenv}, they obey the relation
  \begin{align}
   S_{[4]}\psi_{j_1,j_2,j_3}^{(m)}(\vec{s}) = (m+\gamma_{[4]}-\tfrac{1}{2})\psi_{j_1,j_2,j_3}^{(m)}(\vec{s}).
  \end{align}
  Recalling \eqref{algebraicH}, this also implies
  \begin{align}
   H\psi_{j_1,j_2,j_3}^{(m)}(\vec{s})=(m+\gamma_{[4]})(m+\gamma_{[4]}-2)\psi_{j_1,j_2,j_3}^{(m)}(\vec{s}).
  \end{align}
  Finally, we can normalize these eigenfunctions as
  \begin{align}
  \Psi_{j_1,j_2,j_3}^{(m)}(\vec{s}) = \frac{\eta_1\eta_2\eta_3}{\sqrt{2}}\psi_{j_1,j_2,j_3}^{(m)}(\vec{s}),
  \end{align}
  where
  \begin{align}
   &\eta_1 = \gamma_2\sqrt{\frac{\Gamma(a+\gamma_{[2]})}{a!\Gamma(a+\gamma_1)\Gamma(a+\gamma_2)}}\times
            \begin{cases}
            1 \quad &\text{if $j_1 = 2a$,} \\
            \sqrt{\frac{a+\gamma_2}{a+\gamma_1}} \quad &\text{if $j_1=2a+1$,}
            \end{cases} \\           
   &\eta_2 = \gamma_3\sqrt{\frac{\Gamma(b+j_1+\gamma_{[3]})}{b!\Gamma(b+\gamma_3)\Gamma(b+j_1+\gamma_{[2]})}}\times
            \begin{cases}
            1 \quad &\text{if $j_2 = 2b$,} \\
            \sqrt{\frac{b+\gamma_3}{b+j_1+\gamma_{[2]}}} \quad &\text{if $j_2=2b+1$,}
            \end{cases} \\
   &\eta_3 = \gamma_4\sqrt{\frac{\Gamma(c+j_1+j_2+\gamma_{[4]})}{c!\Gamma(c+\gamma_4)\Gamma(c+j_1+j_2+\gamma_{[3]})}}\times
            \begin{cases}
            1 \quad &\text{if $j_3 = 2c$,} \\
            \sqrt{\frac{c+\gamma_4}{c+j_1+j_2+\gamma_{[3]}}} \quad &\text{if $j_3=2c+1$,}
            \end{cases}           
  \end{align}
  so that 
  \begin{align}
   \int_{S^3} \Psi_{j_1,j_2,j_3}^{(m)\dag}(\vec{s}) \Psi_{k_1,k_2,k_3}^{(n)}(\vec{s}) d\vec{s} = \delta_{n,m}\delta_{j_1,k_1}\delta_{j_2,k_2}.
  \end{align}
  This can be verified directly from the orthogonality relation of the Jacobi polynomials \cite{Askey_scheme}.
\section{Conclusion}
To sum up, we have introduced a new quantum superintegrable model with reflections on the three-sphere. Its symmetries were given explicitly and were shown to realize a rank-two Bannai-Ito algebra. It was observed that the model can be constructed through the combination of four independent realizations of the superalgebra $\mathfrak{osp}(1|2)$. A quadratic expression in the total sCasimir operator was found to coincide with the Hamiltonian while the intermediate Casimir operators were seen to coincide with its symmetries. The exact solutions have been obtained by using a Cauchy-Kovalevskaia extension theorem. We did not find many occurences of this remarkably simple technique in the superintegrability literature and we trust it could find many other applications. Furthermore, an interesting feature of this model is the appearance in a scalar model of the rank 2 Bannai-Ito algebra which arose as a particular case in the analysis of the Dirac-Dunkl equation \cite{higherrankBIalgebra}. One expects that the bivariate Bannai-Ito polynomials will arise as overlaps between wavefunctions of this model separated in different hyperspherical coordinate systems. These polynomials have never been identified so far and we aim to study this question in the near future.    

\section*{Acknowledgments}
The research of HDB is supported by the Fund for Scientific Research-Flanders (FWO-V), project ``Construction of algebra realisations using Dirac-operators'', grant G.0116.13N. VXG holds a postdoctoral fellowship from the Natural Science and Engineering Research Council of Canada (NSERC). JML holds a scholarship from
the Fonds de recherche du Qu\'ebec – Nature et technologies (FRQNT). The research of LV is supported in part by NSERC.



\end{document}